\documentclass[conference]{IEEEtran}
\IEEEoverridecommandlockouts
\usepackage{cite}
\usepackage{amsmath,amssymb,amsfonts}
\usepackage{algorithmic}
\usepackage{graphicx}
\usepackage{textcomp}
\usepackage{xcolor}
\def\BibTeX{{\rm B\kern-.05em{\sc i\kern-.025em b}\kern-.08em
    T\kern-.1667em\lower.7ex\hbox{E}\kern-.125emX}}
    \makeatletter
\newcommand{\linebreakand}{%
  \end{@IEEEauthorhalign}
  \hfill\mbox{}\par
  \mbox{}\hfill\begin{@IEEEauthorhalign}
}
\makeatother

\begin{document}

\title{Characterizing quantum dynamics using multipartite entanglement generation \\
{
}
\thanks{DST SERB Core Research Grant CRG/2021/007095}
}

\author{
\IEEEauthorblockN{Gaurav Rudra Malik}
\IEEEauthorblockA{\textit{Department of Physics} \\
\textit{Indian Institute of Technology (Banaras Hindu University)}\\
Varanasi-221005, India \\
gauravrudramalik.rs.phy22@itbhu.ac.in}   
\and
\IEEEauthorblockN{Rohit Kumar Shukla}
\IEEEauthorblockA{\textit{Center for Quantum Entanglement Science and Technology} \\
\textit{Bar-Ilan University}\\
Ramat-Gan-52900 Israel  \\
rohitkrshukla.rs.phy17@itbhu.ac.in}
\and
\linebreakand
\IEEEauthorblockN{S. Aravinda}
\IEEEauthorblockA{\textit{Department of Physics} \\
\textit{Indian Institute of Technology Tirupati}\\
Tirupati-517619, India \\
aravinda@iittp.ac.in}
\and
\hspace{1.0cm}
\IEEEauthorblockN{Sunil Kumar Mishra}
\IEEEauthorblockA{\hspace{1.0cm} \textit{Department of Physics} \\
\hspace{1.0cm}\textit{Indian Institute of Technology (Banaras Hindu University)}\\
\hspace{1.0cm} Varanasi-221005, India \\
\hspace{1.0cm} sunilkm.app@iitbhu.ac.in}
}

\maketitle

\begin{abstract}
Entanglement is a defining feature of many-body quantum systems and is an essential requirement for quantum computing. It is therefore useful to study physical processes which generate entanglement within a large system, as they maybe replicated for applications involving the said requirements in quantum information processing. A possible avenue to maximize entanglement generation is to rely on the phenomena of information scrambling, i.e. transport of initially localized information throughout the system. Here the rationale is that the spread of information carries with it an inherent capacity of entanglement generation. Scrambling greatly depends upon the dynamical nature of the system Hamiltonian, and the interplay between entanglement generation and information scrambling maybe investigated taking a chain of interacting spins on a one dimensional lattice. This system is analogous to an array of qubits and this relative simplicity implies that the resulting unitary dynamics can be efficiently simulated using present-day cloud based NISQ devices. In our present work, we consider such a spin model which is made up of nearest and next nearest neighbor XXZ Model, along with an introduced coupling term $\lambda$. This coupling term serves as a tuning parameter which modifies the dynamical nature of the system from the integrable to the quantum chaotic regime. In order to quantify the entanglement generated within the system we use the more general multipartite metric which computes the average entanglement across all system bipartitions to obtain a global picture of the entanglement structure within the entire system. Further to make our study initial state independent, we consider an ensemble of random initial states and evaluate the multipartite entanglement generated for each of these states under time evolution. We find clear differences in the entanglement measures as the system undergoes through different regimes. We also analyze the fluctuations present within our entanglement metric and find a distinct pattern as we head from the integrable to the chaotic regime.
\end{abstract}


\section{Introduction}
The study of dynamical models is an active area of research in the field of quantum information and condensed matter physics \cite{k1,k2}. The discussion revolves around the central question of how the wavefunction and its associated properties change when acted upon by a time propagator \cite{k3}. From the perspective of a circuit implementation of the time evolution process, the said propagator is of a unitary type and implemented by using a combination of quantum gates.

One of the primary quantities of interest associated with the time evolution of quantum state is that of entanglement \cite{k4,k5}. This is because entanglement serves as a primary resource for the implementation of nearly all quantum computation/ communication algorithms \cite{k6,k7,k8}. It is, therefore, that a unitary, Hamiltonian based circuit generating high entanglement is useful. On the other hand, when we consider simulating the temporal evolution brought on by a many-body quantum system, it is the low rate of entanglement growth that allows for the applicability of tensor-network based techniques \cite{k9}.

Therefore, the emerging entanglement structure is an important consideration when analyzing the dynamics brought on by a many-body Hamiltonian. It is already well known that the nature of the Hamiltonian of a system has an intimate relation with the entanglement generating nature of the associated quantum circuit \cite{k10,k11}. For example, a non-integrable system displaying quantum chaotic features, in general, always generates greater entanglement across a bipartition than the integrable case \cite{k12}. Another aspect associated with dynamics is the scrambling of information \cite{k13,k14}, which is also helpful in characterizing the nature of the dynamics and differentiating integrable models from chaotic ones. Thus, knowing about non-integrable and chaotic models by their scrambling nature can often help us create quantum circuits which lead to rapid entanglement growth \cite{k15}. This approach of using system dynamics as a generator of entanglement presents an generic, efficient and relatively straighforward approach to generate high entanglement content within a given collection of qubits which is usually a non-trivial problem. 

In current our work we consider an instance of this scenario taking a simple example of a model where the dynamical features are tunable and observing the resulting extent of entanglement arising from the corresponding unitary evolution. Further, we use a metric which extends the notion of entanglement, usually based on the idea of a bipartition within the system to that of multipartite entanglement, and investigate the relation it has with the inherent dynamical nature of the time-evolution. In this context, the tunable parameter plays an important role in modifying the dynamic behaviour shown by the system. For the purpose of calculating our multipartite entanglement metric, we take each part of the system individually and calculate the entanglement with the remaining system, repeating this process for all individual subsystems. In the following sections we discuss our model, define our multipartite entanglement metric along with a way to measure the associated power of the unitary evolution in generating the same. We then provide results which highlight the role of the coupling parameter $\lambda$ in the system dynamics.

This study explores the details of the entanglement structure generated by different classes of increasing dynamical complexity and the results we put forward indicate that increasing the complexity associated with the corresponding time evolution increases the extent of quantum entanglement. Moreover, we introduce another metric that quantifies the oscillatory nature of the entanglement in the long-time limit, which may conversely serve as a discriminator between integrable and chaotic dynamics.

\section{Model}

For our present work, we consider the simplest example of a many-body Hamiltonian, namely the anisotropic Heisenberg model, also called as the XXZ Model. We consider two different versions of the XXZ Hamiltonian, the first involving nearest neighbor (NN) coupling, and the other having next nearest neighbor (NNN) coupling in place of NN coupling. We denote them respectively by the terms $\hat{H}_0$ and $\hat{H}_1$. The mathematical form of the two versions can be written down in terms of the spin vector $S$ and expanded in terms of the individual Pauli matrices acting on the respective sites for a system of size $L$ with open boundary condition:
\begin{align}
    \hat{H}_0 = \frac{J}{2} \sum_{i = 0} ^ {L-1} \hat{S}_j^{x} \hat{S}_{j+1}^{x} + \hat{S}_j^{y} \hat{S}_{j+1}^{y} + \mu \hat{S}_j^{z} \hat{S}_{j+1}^{z} \\
    \hat{H}_1 = \frac{J}{2} \sum_{i = 0} ^ {L-2} \hat{S}_j^{x} \hat{S}_{j+2}^{x} + \hat{S}_j^{y} \hat{S}_{j+2}^{y} + \mu \hat{S}_j^{z} \hat{S}_{j+2}^{z}
\end{align}

Here $\mu$ is termed as the anisotropy factor. The above Hamiltonian has a straightforward representation in the second quantized formalism, involving the mapping between the Pauli and the hard-boson. For that we recast the Pauli operators $\hat{S}_x$ and $\hat{S}_y$ in terms of the raising and lowering operators $\hat{S}_{\pm}$ \cite{k16}.   

In this form, the Hamiltonian with NN coupling reduces to:

\begin{equation}
    \hat{H}_0 = J \sum_{i = 0} ^ {L-1} \hat{S}_j^{+} \hat{S}_{j+1}^{-} + \hat{S}_j^{-} \hat{S}_{j+1}^{+} + \frac{\mu}{2} \hat{S}_j^{z} \hat{S}_{j+1}^{z}
\end{equation}

From here, mapping $\hat{S}_{+}^j$ to $\hat{b}_j^\dagger$ and $\hat{S}_{-}^j$ to $\hat{b}_j$, we get the following form, subject to general bosonic commutation relation and the hard-boson repulsion given as $\hat{b}_j^{\dagger 2} =  \hat{b}_j^2 = 0$. 

\begin{align}
    \hat{H}_0 = J \sum_{i=0}^{L-1} \hat{b}_j^\dagger \hat{b}_{j+1} + \hat{b}_j \hat{b}_{j+1}^\dagger + \frac{\mu}{2} (2\hat{n}_j - 1)(2\hat{n}_{j+1} - 1); \\
    \hat{H}_1 = J \sum_{i=0}^{L-2} \hat{b}_j^\dagger \hat{b}_{j+2} + \hat{b}_j \hat{b}_{j+2}^\dagger + \frac{\mu}{2} (2\hat{n}_j - 1)(2\hat{n}_{j+2} - 1);
\end{align}

for the NN and the NNN versions of the Hamiltonian. Here $\hat{n}_j = \hat{b}_j^\dagger \hat{b}$ is the number operator, and we have used the mapping $\hat{S}_j^{z} = 2\hat{n}_j - 1$ in the final terms of the expression.

From here, the given Hamiltonian may be solved analytically using the Bethe ansatz. However, the action of this Hamiltonian on a state $|n_{L-1}\rangle \otimes |n_{L-2}\rangle ... \otimes |n_0\rangle$ of the associated hilbert state is apparent to observe. The term involving the factor $\mu$ is diagonal in its action, and adds a multiplicative factor to the state it is operating on. The remaining part acts as a flip operator, exchanging the $j th$ and $(j+1) th$ ($(j+2)th$) entries of the $L$ length bit string defining the input state for the NN (NNN) model.

Here, it is important to observe that in the absence of anisotropy (i.e. $\mu = 1$), the Hamiltonian has the effect of simply applying a series of SWAP gates, which does not lead to any entanglement generation \cite{k17,k18}. While the Hamiltonian terms $\hat{H}_0$ and $\hat{H}_1$ are straightforward in their action, for our work we consider a system having both these terms along with a coupling parameter $\lambda$, with $\mu \ne 1$. Thus, our system Hamiltonian is defined as:

\begin{align}
    \hat{H}_{sys} = \hat{H}_0 + \lambda \hat{H}_1
\end{align}

Given the nature of the hopping terms for the components $\hat{H}_1$ and $\hat{H}_2$, it is apparent that an element of frustration also exists in the model, which is dependent on the coupling term $\lambda$, and is maximum when $\lambda \approx 1$. The model also transitions from integrable to chaotic dynamics on varying the same parameter \cite{k23}. This is a key feature of our model, and based on this we carry out our investigations analyzing the role of the dynamical nature with the multipartite entanglement generated within the system. In the following sections we shall study the entanglement generation as a result of the unitary evolution arising due to the above Hamiltonian, investigating primarily the effect of the coupling parameter $\lambda$ as it governs the dynamical nature of the system. This model, and the entanglement metric may also be easily implemented as a quantum circuit, and run of a real cloud-based NISQ device for a digital quantum simulation.

\section{Entanglement Metrics}

Entanglement across a bi-partition for a pure quantum state is well-defined for any two arbitrary subsystems $A$ and $B$ within the larger system. It is quantified using the von-Neumann entropy of partial density matrices or more commonly by the linearized version of the same, given by:

\begin{equation}
    S_L(\psi) = \eta \Big( 1 - tr\rho_{A}^2 \Big)
\end{equation}

where $\rho_A = tr_B(|\psi\rangle \langle \psi|)$. It represents the degree of mixedness within the subsystem when traced out from an entangled state $|\psi\rangle \in \mathbb{C}^{d_A} \otimes \mathbb{C}^{d_B}$. $\eta$ is the associated normalization factor, and is given by $1/d(d+1)$ where $d = min(d_A,d_B)$. We have $S_L(\psi) \in [0,1]$, and the limits are reached for separable and maximally entangled states respectively.

As the system size increases, there is a rapid increase in the number of possible bi-partitions, each having a corresponding measure of entanglement. An average of the entanglement across all such bi-partitions forms a simple measure of the multipartite nature of entanglement. Indeed, the Mayer-Wallach measure \cite{k19} of multipartite entanglement $Q(\psi)$ defined for a multiqubit state can be written in exactly such a form \cite{k20}:

\begin{equation}
    Q(\psi) = 2 \Big( 1 - \frac{1}{L} \sum_{k = 1}^{L} tr\rho_k^2 \Big)
\end{equation}

where $\rho_k$ is the density matrix corresponding to the $k$th spin, when all the remaining spins have been traced out. In effect, it represents the average entanglement of each spin with the remaining system. A more general form of this is also possible, taking all possible $m$ sized clusters and evaluating their average entanglement with the remaining system, known as the Scott Measure $Q^m(\psi)$ \cite{k21}. However for our purposes the Mayer-Wallach measure is good enough, and maybe referred to as $Q^{m=1}(\psi)$ following the notation of the general Scott Measure.

With this measure, given a unitary operator $U$ acting upon a given initial state, an average of the measure $Q(\psi)$ over an ensemble of uniformly distributed states indicates the ability of the operator $U$ to generate multipartite entanglement. We refer to this quantity by $\mathbb{E}(U)$ and write its explicit mathematical form using the Haar Random ensemble \cite{k22} as follows:

\begin{equation}
    \mathbb{E}(U) = \int d\mu(\psi_1,\psi_2 ... \psi_L) Q(U |\psi_1\rangle \otimes |\psi_2\rangle \otimes ... |\psi_L\rangle)
\end{equation}

This is a more fundamental measure to analyze the ability of a circuit to generate correlations within the system, as it is both initial-state independent, and also local unitary invariant owing to the inherent properties of the Haar-Random distribution. In the following sections we shall put forward results for the variation of $\mathbb{E}(U(t))$ where $U(t) = e^{(-i\hat{H}_{sys}t)}$, with time. This shows distinct variations for different dynamical regimes of the $\hat{H}_{sys}$ Hamiltonian.    

\begin{figure}
    \centering
    \includegraphics[width= 0.8\linewidth, height = 0.7\linewidth]{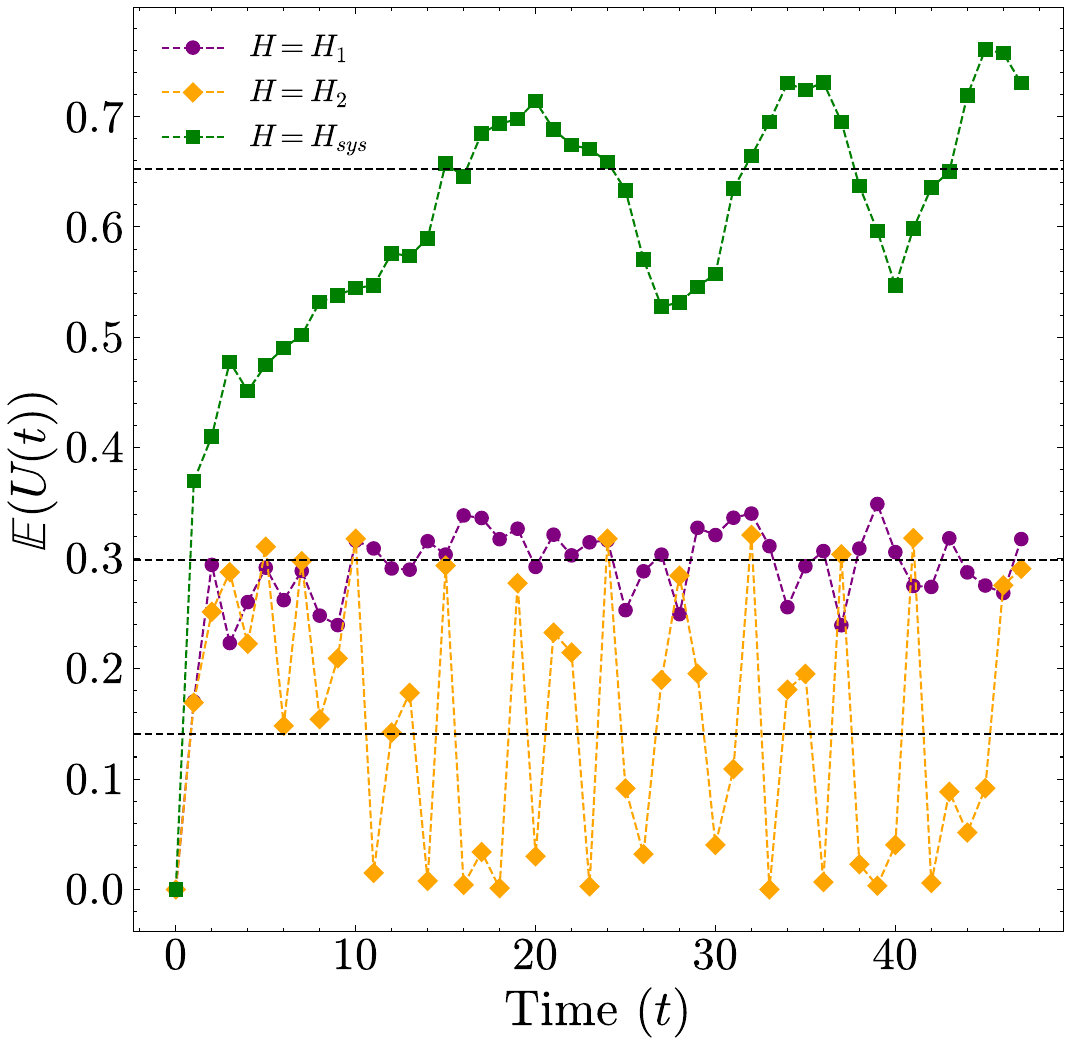}
    \caption{Average entanglement generation for NN Hamiltonian $(\hat{H}_1)$, NNN Hamiltonian $(\hat{H}_2)$ and $\hat{H}_{sys}$ with $\lambda = 1$. $\mu = 1.5$ and $N = 8$ for all cases, over an ensemble of 100  Haar Random samples. }
    \label{figure1}
\end{figure}

\section{Results and Discussions}

We begin by plotting the average multipartite entanglement generated by different components of our Hamiltonian i.e. $\hat{H}_1$ and $\hat{H}_2$ involving the NN and NNN interactions. As observed in the Fig.\ref{figure1} the correlation generated by $\hat{H}_{sys}$ is much higher than those generated by either $\hat{H}_{1/2}$. Moreover, the entanglement produced by $\hat{H}_{1/2}$ has an entirely oscillatory character, which is in contrast to that observed for $\hat{H}_{sys}$ which has an on average increasing trend throughout the time of observation. Here we also mention that the manner in which we are constructing the initial Haar-Random state is such that it has zero initial entanglement. Also we have $\mu = 1.5$ for all cases. For the case where $\mu = 1.0$, we shall not see an increase in entanglement. This is because at $\mu = 1$ the anisotropy term becomes trivial, as is apparent by Eq. 4 and 5. In this case the remaining hopping terms just act as SWAP operators which do not have any entanglement generating capacity \cite{k17,k18}.

It is also apparent the as the average entanglement increases, the expected entanglement, which we define as the time average of the generated multipartite entanglement, also sees a rise. This is indicated by the straight lines in the Fig. \ref{figure1} which rises from a value of $1.5$ for $\hat{H}_2$ to $3.0$ for $\hat{H}_1$ to finally $6.5$ for $\hat{H}_{sys}$. It therefore becomes clear the that $\hat{H}_{sys}$ generates greater inter-system correlation that any of its two components. This is a remarkable result, because it serves as an example where the coupling between two integrable models generates dynamical features beyond their individual capabilities. This is also a clear indication of inherent chaotic features underlining the time evolution generated by $\hat{H}_{sys}$. This results aligns with previous work investigating the system $\hat{H}_{sys}$ where it was concluded that the model becomes chaotic in the regime when $\lambda \approx 1$ \cite{k23}. Moreover, it suggests a computationally efficient manner of generating chaotic features using two integrable and classically tractable models involving the general Heisenberg exchange interaction.

\begin{figure}
    \centering
    \includegraphics[width=0.8\linewidth, height = 1.5\linewidth]{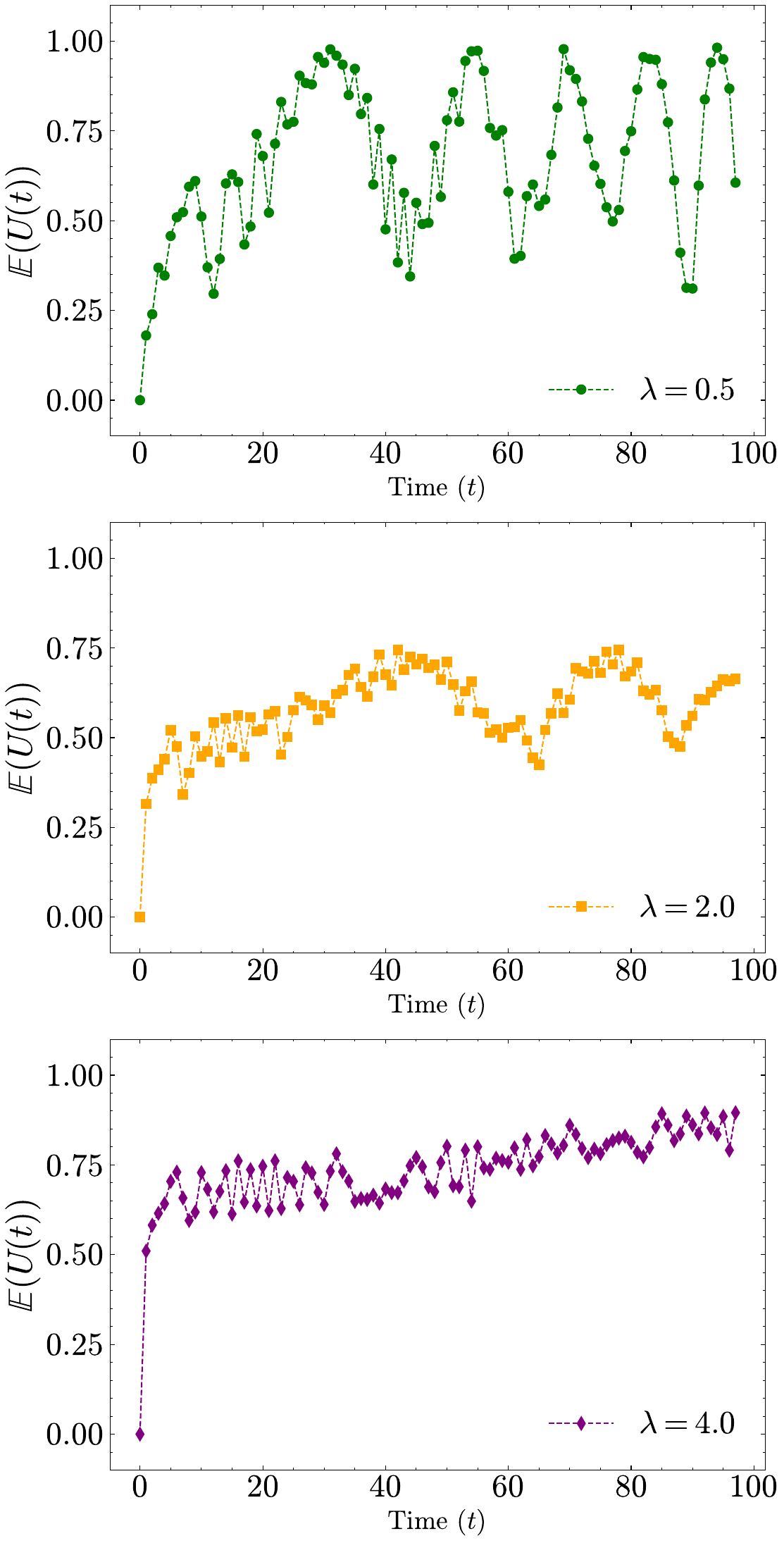}
    \caption{Average entanglement generation for $\hat{H}_{sys}$ having $\lambda = 0.5$, $\lambda = 2.0$ and $\lambda = 4.0$. $\mu = 1.5$ and $N = 8$ for all cases, over an ensemble of 100  Haar Random samples.}
    \label{figure2}
\end{figure}

More indications of this transition from integrable to chaotic can be found by individually plotting the average multipartite entanglement generated for differing values of $\lambda$. Results related to this transition are shown in Fig. \ref{figure2}. For $\lambda = 0.5$, which corresponds to the integrable regime, we see a gradual increase $\mathbb{E}(U^t)$ as we increase the value of time $t$. This gradual increase settles down to a fixed value, superimposed by a distinctly oscillatory behavior. Comparing this to $\lambda = 2.0$ and $\lambda = 4.0$ we can see that the initial growth of the average multipartite entanglement is abrupt, and takes just a few steps ($\approx$ 5) to reach a high value. Moreover there is a clearly visible linear growth in $\mathbb{E}(U^t)$ with time which is distinct to the fixed expected entanglement as observed for the integrable case. Most apparent however, is the diminishing oscillatory behavior as we increase the value of $\lambda$. This is the strongest indicator that the dynamics have transitioned from an integrable to a chaotic regime \cite{k24, k25}. This is a behavior which is commonly observed in several metrics related to the scrambling of information, most notable in the out of time order correlation (OTOC). The physical reason behind this observation is as follows: For an integrable systems there are always associated quantities which remain conserved throughout the dynamics. This restricts the accessible Hilbert space for the system to propagate in and causes the repetitive cycles which are observed as oscillations. No such restriction exists for the chaotic case. This enables the system to propagate towards a unilateral direction, which is often combined with a increase in correlations amongst all the constituents present within. Therefore analyzing the present oscillations maybe helpful in discriminating chaotic dynamics from instances of regular/integrable dynamical features.

\begin{figure}
    \centering
    \includegraphics[width=0.8\linewidth, height=0.7\linewidth]{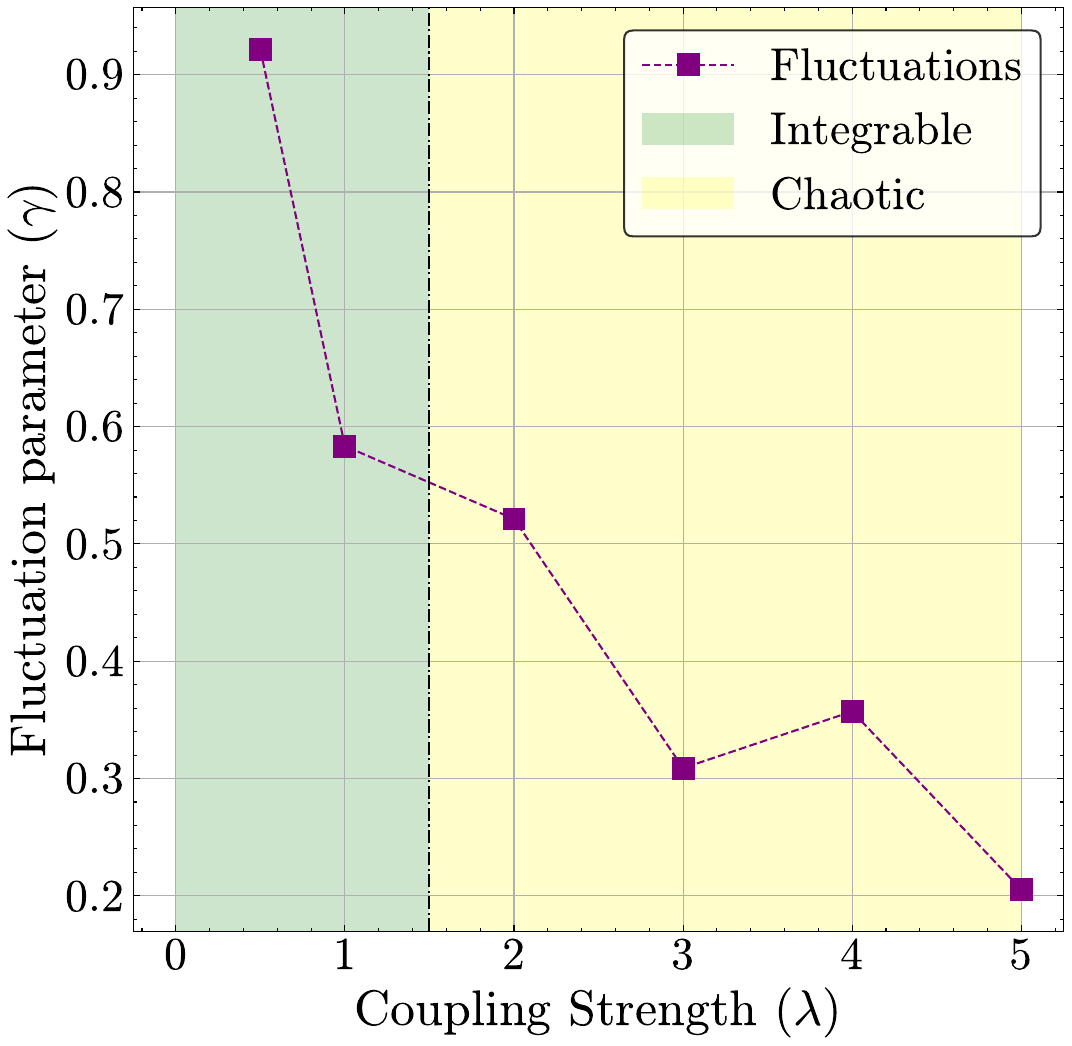}
    \caption{Variation of $\gamma$ with the coupling parameter $\lambda$. Here $N = 8$ and $\mu = 1.5$ for all the different values of $\lambda$, over an ensemble of $100$ Haar Random samples.}
    \label{figure3}
\end{figure}

We finally present a metric for characterizing the oscillations in order to see the effect of increasing $\lambda$. This is defined as:
\begin{equation}
    \gamma = \frac{max(\mathbb{E}(U(t))) - min(\mathbb{E}(U(t)))|_{t \in [\tau,T_{max}]}}{\frac{1}{T_{max} - \tau} \int_{\tau} ^ {T_{max}} \mathbb{E}(U(t)) dt}
\end{equation}

It is effectively a ratio of the range of average multipartite entanglement between time $t = \tau$ and $t = T_{max}$ and the expected entanglement value. For our case, $T_{max} = 100$, and we take $\tau = 20$, which is the point after which the multipartite entanglement generation appears to stabilize. Observing the results in Fig.\ref{figure3} we can clearly see the transition from integrable to chaotic dynamics upon increasing the value of $\lambda$. It maybe noted that for our model, increasing the parameter $\lambda$ increases the frustration present in the model, which is maximized at $\lambda \approx 1$. This raises the interesting prospect of investigating frustration as a generator of quantum chaos with spin models, particularly for the $2D$ case where other physical features causing geometrical frustration are also well studied. A common example of this includes the antisymmetric exchange on a $2D$ lattice, resulting from the Dzyaloshinskii–Moriya interaction which induces non-colinearity within spins present in a regular lattice.

\section{Conclusion}

In our present work, we consider a Hamiltonian that is obtained by combining via a coupling factor $\lambda$, two versions of the 1D integrable Heisenberg model, involving nearest-neighbor and next nearest-neighbor interactions. It is known that dependent on $\lambda$ the model shows integrable ($\lambda \le 1$) and chaotic behavior ($\lambda > 1$). We investigate the system in these two regimes by analyzing the multipartite entanglement generation for both of these cases. Being an example of non-classical correlation amongst all possible segments of the system, multipartite entanglement should be effected by the dynamics and rate of scrambling, and this is found to be affirmative for our particular case. It is observed that for the integrable regime the growth of average multipartite entanglement over an ensemble of Haar-Random initial states is gradual, and also towards the later stage consists of a distinct oscillations. This is in contrast to the observations made for the chaotic regime, where the average multipartite entanglement abruptly reaches a high value, and instead of distinct oscillations we observe a persistent trend of linear growth along with severely subdued fluctuations. Although similar observations we made for different metrics such as the OTOCs, with our measure, we directly link the correlations within all subparts of the system with the dynamics it undergoes, using our metric of multipartite entanglement. We find most remarkable of all, the ease with which we can observe chaotic features combining together two different versions of a model with very simple dynamics of its own. It further provides an example where frustration in spin models is a driver of chaotic dynamics, and this is an area which can be explored in other research projects. Further, since the model is ultimately based on the Heisenberg chain, an implementation of the associated time evolution can be easily implemented using a NISQ computer for higher system sizes. Finally, our Hamiltonian $\hat{H}_{sys}$ generates high values of multipartite entanglement, if allowed to act on the system for a long time, providing us with a technique for the same. States with high multipartite entanglement, although non-trivial to prepare, are arguably more useful for implementing practical applications owing to the higher number of subsystems being involved in the entanglement structure. 

\section*{Acknowledgment}

We gratefully acknowledge the support provided by the project ``Study of quantum chaos and multipartite entanglement using quantum circuits'' sponsored by the Science and Engineering Research Board (SERB), Department of Science and Technology (DST), India under the Core Research Grant CRG/2021/007095. Further, GRM acknowledges the support from Student Travel Grant Support, IIT(BHU) for presenting this work at the COMSNETS 2025 conference. We are also grateful to the referees for providing valuable comments that have improved upon the content and presentation of the original manuscript



\begin{thebibliography}{00}
\bibitem{k1} E. Iyoda and T. Sagawa, Phys. Review A 97, 042330 (2018)
\bibitem{k2} M. Fisher, V. Khemani, A. Nahum and S. B.Vijay, Ann. Rev. Con.Mat.Phys, Vol. 14:335-379 (2023) 
\bibitem{k3} B. Swingle, G. Bentsen, M. Schleier-Smith, and P. Hayden, Phys. Rev. A 94, 040302 (2016)
\bibitem{k4} S. K. Mishra, A. Lakshminarayan, and V. Subrahmanyam, Phys. Rev. A 91, 022318 (2015)
\bibitem{k5} G. K. Naik, R. Singh, and S. K. Mishra, Phys. Rev. A 99, 032321 (2019)
\bibitem{k6} A. K. Ekert, Phys. Rev. Lett. 67, 661 (1991)
\bibitem{k7} C. H. Bennett and S. J. Wiesner, Phys. Rev. Lett. 69, 2881 (1992)
\bibitem{k8} C. H. Bennett, G. Brassard, C. Cr´epeau, R. Jozsa, A. Peres, and W. K. Wootters, Phys. Rev. Lett. 70, 1895 (1993)
\bibitem{k9} J. Eisert, M. Cramer, and M. B. Plenio, Rev. Mod. Phys. 82, 277 (2010)
\bibitem{k10} T. Prosen, J. Phys. A: Math. Theor. 40 7881 (2007)
\bibitem{k11} X. Wang, S. Ghose, B. Sanders, and B. Hu Phys. Rev. E 70, 016217 (2004)
\bibitem{k12} R. Pal and A. Lakshminarayan, Phys. Rev. B 98, 174304 (2018)
\bibitem{k13} R. K. Shukla, arXiv preprint arXiv:2310.14620 (2023)
\bibitem{k14} Y. Kuno, T. Orito, and I. Ichinose, Phys. Rev. A 106, 012435 (2022)
\bibitem{k15} P. W. Claeys and A. Lamacraft, Phys. Rev. Research 2, 033032 (2020)
\bibitem{k16} J. H. Jung and J. D. Noh, Journal of the Korean Physical Society, Volume 76 (2020)
\bibitem{k17} B. Jonnadula, P. Mandayam, K. Życzkowski, and A. Lakshminarayan, Phys. Rev. Research 2, 043126 (2020)
\bibitem{k18} S. Aravinda, S. A. Rather, and A. Lakshminarayan, Phys. Rev. Research 3, 043034 (2021)
\bibitem{k19} D. A. Meyer and N. R. Wallach, J. Math. Phys. 43, 4273 (2002)
\bibitem{k20} G. K. Brennen, Quantum Inf. Comput. 3, 619 (2003)
\bibitem{k21} A.J. Scott, Phys. Rev. A 69, 052330 (2004)
\bibitem{k22} A.A. Mele, Quantum, Volume 8 (2024)
\bibitem{k23} L. F. Santos, F. Borgonovi, and F. M. Izrailev, Phys. Rev. E 85, 036209 (2012)
\bibitem{k24} E. M. Fortes, I. García-Mata, R. A. Jalabert, and D. A. Wisniacki, Phys. Rev. E 100, 042201 (2019)
\bibitem{k25} R. K. Shukla, G. R. Malik, S. Aravinda, S. K. Mishra, arXiv:2404.04177 [quant-ph]

\end{thebibliography}
\end{document}